# Prospects for measuring Higgs properties at the LHC

H. Przysiezniak

*LAPP, 9 Chemin de Bellevue, BP 110, F-74941 Annecy-le-Vieux CEDEX, France*

In the following, the prospects for measuring the Standard Model (SM) Higgs properties at the LHC (Large Hadron Collider) are reviewed, in particular its mass, width, spin, CP quantum numbers as well as its couplings to the SM fermions and gauge bosons. The possibility of performing the difficult trilinear Higgs self-couplings measurement is also discussed.

## 1 Introduction

The discovery of new particles at the LHC will depend on nature itself, but also on the readiness of the LHC machine and of our detectors e.g. ATLAS (A Toroidal LHC Apparatus) and CMS (Compact Muons Solenoid). In the coming months before collisions occur within the LHC (expected for 2007), the experiments will need to commission their detectors and associated electronics, as well as trigger systems. New physics will be potentially accessible once enough integrated luminosity is accumulated e.g. 10 fb$^{-1}$ or less per experiment are necessary for a SM Higgs discovery (signal significance S/$\sqrt{B} > 5\sigma$) depending on its mass.

If some Higgs-like particle is discovered, one will want to measure its properties, some more straightforward than others: mass, width, spin, CP quantum numbers, couplings to SM fermions and gauge bosons, as well as self-couplings. The SM Higgs boson is produced via gluon fusion (GF), followed by weak vector boson fusion (WBF), and finally in associated production (W, $t\bar{t}$, Z and $b\bar{b}$). For a Higgs heavier than the LEP exclusion limit of 114.4 GeV/$c^2$, the decay modes of interest are: two-photon in GF and WBF, $\tau\tau$ in WBF, as well as WW and ZZ decays.

## 2 Mass and width

The mass is extracted directly by measuring the position of the peak in the invariant mass distribution, for the following channels: H $\to \gamma\gamma$, $t\bar{t}$, W(H $\to b\bar{b}$), H $\to$ ZZ$^{(*)} \to 4\ell$, as well as WBF H $\to \tau\tau \to \ell+$ hadrons, where one of the $\tau$ leptons decays leptonically and the other hadronically. The mass can also be extracted indirectly, by comparing various variable distributions, which are sensitive to the Higgs mass, to Monte Carlo (MC) templates. This can be done for the following channels: H $\to$ WW $\to \ell\nu\ell\nu$, W($H \to WW$) $\to \ell\nu(\ell\nu\ell\nu)$, and WBF H $\to \tau\tau \to \ell\ell$.

Using 300 fb$^{-1}$ of data while combining the expected results of the CMS and ATLAS experiments (see e.g. [1] for the ATLAS results), a relative precision on $m_H$ of 0.1% can be obtained for $m_H = 100 - 400$ GeV/$c^2$. From Figure 1, one can see that for $m_H > 400$ GeV/$c^2$, the precision degrades down to 1% at $m_H = 700$ GeV/$c^2$. The systematic uncertainties are dominated by the knowledge of the absolute calorimetric energy scale. For leptons and photons, it has been assumed to be 0.1%, but the absolute goal is 0.02%, namely in order to attain an error of 15 MeV on the W mass. For jets, a 1% energy scale uncertainty is assumed.

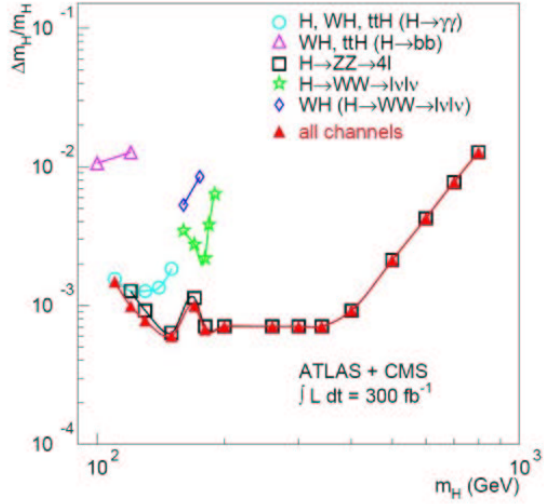

Figure 1: Relative error on $m_H$ for CMS and ATLAS combined assuming 300 fb$^{-1}$ of data.

In what concerns the Higgs width, a direct measurement is only possible for $m_H > 200$ GeV/$c^2$, when the Higgs width becomes comparable or larger than the intrinsic detector resolution ($\sim$1 GeV) e.g. $\Gamma_H \sim 3$ MeV for $m_H = 110$ GeV/$c^2$ and $\Gamma_H \sim 1$ GeV for $m_H = 200$ GeV/$c^2$. In this case, a relative precision of 6% can be attained. The indirect extraction method will be described in Section 4.

## 3 Spin and CP eigenvalues

If a Higgs-like particle is observed, one will want to know if it has spin and CP eigenvalues $J^{CP} = 0^{++}$. An ATLAS analysis [2] has studied the possibility to use angular distributions and correlations in the H $\to$ ZZ $\to 4\ell$ channel for $m_H > 200$ GeV/$c^2$, in order to extract the (J,CP) state of the resonance. Two angular distributions are investigated:

- $\cos\theta$ : the polar angle of the leptons relative to the Z boson in the Higgs rest frame;
- $\phi$ : the angle between the decay planes of the two Z bosons in the Higgs rest frame.

The measured distributions are compared to those of hypothetical particles with the following (J,CP) values : (0,1) which is the SM scalar case, as well as (0,-1), (1,1) and (1,-1), respectively the pseudo scalar, the vector and axial vector cases. From these comparisons, one can extract the significance for having a SM Higgs.

For 100 fb$^{-1}$, the polar angle distribution enables to exclude non-SM (J,CP) values for $m_H > 250$ GeV/$c^2$. Furthermore, for 300 fb$^{-1}$ and $m_H > 200$ GeV/$c^2$, (J,CP)=(1,1) and (1,-1) can be ruled out with respectively 6.4$\sigma$ and 3.9$\sigma$ significances. The systematic uncertainties on these results are dominated by background subtraction. For lower Higgs masses ($m_H < 200$ GeV/$c^2$), the azimuthal separation of leptons in WBF production followed by the H $\to$ WW $\to \ell\nu\ell\nu$ decay can be exploited to extract significances on the spin and CP eigenvalues, but this is still work in progress. In addition, the mere observation of non-zero H$\gamma\gamma$ and Hgg couplings rules out J=1 particles and all odd spin particles in general [3].

## 4 Couplings

By measuring the rates for various Higgs production and decay channels, couplings or coupling ratios can be extracted. The coupling constants for weak vector bosons and for fermions are

given by:

$$g_{W,Z} = 2m_{W,Z}^2/v \quad \text{and} \quad |g_f| = \sqrt{2}m_f/v$$

where $v = (\sqrt{2}G_F)^{-1/2} \sim 246$ GeV is the vacuum expectation value of the Higgs field. A first study performed by ATLAS[4] combines channels using a maximum likelihood method for $110 < m_H < 190$ GeV/$c^2$ which enables to extract $g_W$, $g_Z$, $g_t$, $g_b$ and $g_\tau$. For all signal channels, the following can be written in the case of the narrow width approximation:

$$\sigma_H \times \text{BR}(H \to yy)_i(x) = \left(\frac{\sigma_H^{SM}}{\Gamma_{\text{prod}}^{SM}}\right)\left(\frac{\Gamma_{\text{prod}}\Gamma_{H \to yy}}{\Gamma^{\text{total}}}\right)$$

where $x$ is the vector containing the Higgs couplings parameters as well as the quantities with systematic uncertainties e.g. luminosity, detector effects, theoretical uncertainties, etc. The background term $\sigma_H \times \text{BR}(x)$ is treated as a systematic uncertainty. The signal channels considered are: GF production followed by $H \to WW, ZZ, \gamma\gamma, \tau\tau$ ($2\ell$ or $1\ell + 1hadr.$: either the two $\tau$s decay leptonically, or one leptonically and the other hadronically); $t\bar{t}H$ associated production with $H \to WW$ and $t \to Wb$ ($3\ell + 1hadr.$ or $2\ell + 2hadr.$), $H \to \gamma\gamma$, $H \to b\bar{b}$; WH associated production with $H \to WW$ ($3\ell$ or $2\ell + 1hadr.$), $H \to \gamma\gamma$; and ZH associated production with $H \to \gamma\gamma$. As well, the dominant associated backgrounds are considered.

In this study, the following progressive assumptions are made:

- The observed Higgs particle is assumed to be CP-even and spin-0.
  There can be more than one Higgs, but they all have to be degenerate in mass. In this case, only rate measurements are possible.

- There is only one Higgs.
  Any additional Higgs is separated in mass and may not contribute to the channels considered here. The ratio $BR(H \to XX)/BR(H \to WW)$ is equivalent to $\Gamma_X/\Gamma_W$.

- Only the dominant SM couplings prevail.
  There are no extra particles or extremely strong couplings to light fermions. In this case, the squared ratios of the Higgs couplings $g_X^2/g_W^2$ can be extracted, and a lower limit on $\Gamma_H$ is obtained from the sum of all visible decay modes.

- The sum of all visible branching ratios (BRs) is approximately equal to the SM sum.
  In this case, the absolute coupling values can be extracted, as well as the total Higgs width.

Here the last case is described, and $\Gamma_H$ is fixed assuming that the fraction of non detectable Higgs decay modes is as small as in the SM.

The relative error on the squared couplings and for the total width measurements are shown in Figure 2. Assuming 300 fb$^{-1}$ and for $110 < m_H < 150$ GeV/$c^2$, the squared couplings can be extracted with a relative precision of 25% to 60%, except for $g^2(H, b)$ where the relative error goes from 80% to 300%, suffering from the fact that the $H \to b\bar{b}$ channel is difficult to detect. The total width is estimated with a relative error of 50% to 75%. For $160 < m_H < 190$ GeV/$c^2$, the squared coupling measurements profit from the high BRs of $H \to WW, ZZ$, and resolutions of 10% to 15% are attained. The same relative error is obtained for the total width measurement. The main systematic uncertainties originate from efficiency measurements, background normalization and cross sections, as well as parton distribution functions (PDFs).

In another study[5], only one assumption is made, namely that the strength of the Higgs couplings to weak bosons does not exceed the SM value: $\Gamma_V \leq \Gamma_V^{SM}$ for $V = W, Z$. This is justified in any model with an arbitrary number of Higgs doublets and is sufficient to allow the extraction of absolute couplings as well as of the total Higgs width. Assuming 300 fb$^{-1}$ and for $110 < m_H < 190$ GeV/$c^2$, a relative error of 10% to 45% can be obtained on all squared couplings except for $g^2(H, b)$, and 10% to 50% on the total width.

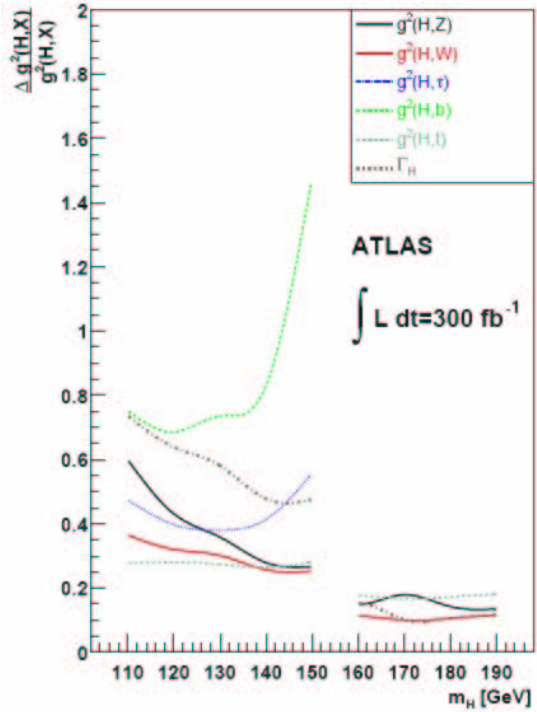

Figure 2: Relative error on $g^2$ and on $\Gamma_H$. The discontinuity at $m_H$ =150-160 GeV/$c^2$ originates from the change in assumption for the sum of all BRs. For $m_H < 150$ GeV/$c^2$, the BRs into b quarks and $\tau$ leptons are included, while above 150 GeV/$c^2$, they are not added to the sum.

## 5 Self-coupling

To establish the Higgs mechanism experimentally, one must reconstruct the Higgs potential given as a function of the physical Higgs boson:

$$V = (m_\mathrm{H}^2/2)H^2 + (m_\mathrm{H}^2/2v)H^3 + (m_\mathrm{H}^2/8v^2)H^4,$$

and measure the trilinear and quadrilinear Higgs self-couplings uniquely determined by $m_\mathrm{H} = \sqrt{2\lambda}v$. One can consider the latter a rather hopeless case at the LHC. Hence only the trilinear self-coupling measurement is investigated here. In the theoretical study performed in[6], the same sign dilepton final state from the following channel

$$gg \to HH \to (W^+W^-)(W^+W^-) \to (jj\ell^\pm\nu)(jj\ell^\pm\nu)$$

is used, for $150 < m_\mathrm{H} < 200$ GeV/$c^2$. It is the only channel which is not swamped by background or which does not have a vanishing cross section. The final states of the main backgrounds are : WWWjj (where j=jets) and $t\bar{t}$W, but also WWjjjj, WZjjjj, $t\bar{t}$Z, $t\bar{t}$j, $t\bar{t}t\bar{t}$, WWWW, WWZjj, overlapping events and double parton scattering. Table 1 lists the signal and background cross sections (in fb) after selection cuts as a function of the Higgs mass. At best, 50 signal events or less will be left after cuts assuming 300 fb$^{-1}$.

The visible system invariant mass $m_{vis}$ distribution is used to extract the value of $\lambda = m_\mathrm{H}^2/2v^2$. Two extra samples with non-SM values of $\lambda_{HHH} = \lambda/\lambda_{SM}$ (0 and 2) are generated. The 95% CL (confidence level) bounds are derived from a $\chi^2$ fit to $m_{vis}$. It is assumed that the SM is valid except of course for the self-coupling, that $m_\mathrm{H}$ is precisely known and that BR(H→WW) is known to 10% or better. Assuming 300 fb$^{-1}$, $\Delta\lambda_{HHH} = (\lambda - \lambda_{SM})/\lambda_{SM} = -1$ (vanishing self-coupling) is excluded at 95% CL or better, and $\lambda$ can be determined with a relative error of -60% and +200%, respectively the lower and upper $1\sigma$ errors. With the same integrated luminosity, the significance for a SM signal is greater than $1\sigma$ for $150 < m_\mathrm{H} < 200$ GeV/$c^2$, and approximately $2.5\sigma$ for $160 < m_\mathrm{H} < 180$ GeV/$c^2$.

Table 1: Cross sections (in fb) after selection cuts for the signal (gg→HH same sign dilepton final state) and summed backgrounds, as a function of $m_\mathrm{H}$.

| $m_\mathrm{H}$ (GeV/$c^2$) | $\sigma$ HH signal (fb) | $\sigma$ total background (fb) |
|---|---|---|
| 150 | 0.07 | 0.90 |
| 160 | 0.19 | 1.03 |
| 180 | 0.18 | 0.94 |
| 200 | 0.08 | 0.83 |

## 6  Conclusion

If a Higgs-like particle is discovered at the LHC, many of its properties can be investigated. Its mass can be measured with a precision of 0.1% to 1% over the whole mass range. A spin value of J=1 would be ruled out assuming 100 (300) fb$^{-1}$ for $m_\mathrm{H}$ >230 (200) GeV/$c^2$, while (J,CP)=(0,-1) could be ruled out with less than 100 fb$^{-1}$ and for $m_\mathrm{H}$ >200 GeV/$c^2$. The squared couplings could be measured with a precision of 10% to 45% (except for $g^2(\mathrm{H,b})$), and in the context of this analysis, the total width would come out with a relative error of 10% to 50%. Finally, one could still say something about the difficult self-coupling measurement: assuming 300 fb$^{-1}$, a vanishing self-coupling could be excluded and $\lambda$ determined to (-60%,+200%).

### Acknowledgments

I would like to warmly thank the organizers of the Rencontres de Moriond conference.